\documentclass[12pt]{article}
\usepackage{textcomp}

\usepackage{graphicx}
\usepackage{xcolor}

\usepackage{amsmath}
\usepackage{amssymb}
\usepackage[natbibapa]{apacite}

\usepackage{array,booktabs}
\usepackage{float}
\usepackage{subcaption}
\usepackage{threeparttable}
\usepackage{url}

\title{Enhancing the Accuracy of Regional Input--Output Table Estimation: A Deep Learning Approach}
\author{Shogo Fukui \thanks{Department of Economics, Seinan Gakuin University, Japan. \\ \hspace*{15px} E-mail: sfukui@seinan-gu.ac.jp}}
\date{}

\begin{document}

\maketitle

%%% abstract
\begin{abstract}
Non-survey methods have been developed for estimating regional input--output tables; however, there is ongoing debate about the necessary assumptions and their accuracy.

To address these issues, this study presents a deep learning method for estimating regional input--output tables.
First, regional quantitative economic data are augmented by linear combinations.
Deep learning is then applied to the items in the input--output table, treating each item as the target variable.
Finally, regional input--output tables are estimated through matrix balancing based on the predicted values from the trained model.

The estimation accuracy of this method is evaluated using the 2015 input--output table for Japan as a benchmark.
Compared with matrix balancing under the ideal assumption of known row and column sums, our method generally demonstrates higher estimation accuracy.
Thus, this method is anticipated to provide a foundation for more precise estimates of regional input--output tables.
\end{abstract}

\vspace*{1ex}

\section{Introduction}
In macroeconomic quantitative analysis of regional economies, input--output tables are fundamental datasets.
% \citep{Miller2022}
Creating these tables requires a large number of quantitative attributes about the region.
Regions with limited data face significant challenges in their construction.
Consequently, considerable research has focused on developing and improving estimation methods for regional input--output tables.

An input--output table records product flows within a target region from two perspectives: producers (input) and consumers (demand) \citep{Leontief1966_1}.
In general, input--output tables are compiled at the national level, while those for domestic regions are referred to as regional input--output tables.

Regional input--output tables are used in analyses of regional economies.
For example, these tables are utilized to analyze the effects of environmental policies and natural disasters on the regional economy.
\cite{Okuyama2007} reviewed studies on disaster impacts on regional economies, many of which employed regional input--output tables.
\cite{Koks2016} evaluated disaster effects on regions using various methods based on regional input--output tables, comparing these approaches.
\cite{Kjaer2015} calculated the environmental footprint using Denmark's environmental input--output table.

However, regional input--output tables are rarely compiled and published for small domestic regions.
In Japan, the national input--output table is compiled from primary statistics \citep{MIC2021_1}.
\footnote{Within the administrative divisions, Japan comprises 47 prefectures, each containing municipalities, such as cities, towns, and villages.}
Prefectural input--output tables are estimated by combining available partial primary data with data from original surveys.
At the municipal level, the availability of primary statistics is even more limited, resulting in very few municipalities publishing their own tables.

In light of this, numerous methods have been developed to estimate regional input--output tables.
Particular focus has been placed on non-survey methods that estimate input--output tables from published statistics without conducting original surveys in the target region.

Most non-survey methods are designed to estimate input coefficients.
Input coefficients, defined as the ratio of intermediate input to the gross output of an input industry, are crucial for obtaining the entire input--output table and calculating economic effects \citep{Leontief1966_1, Lahr2001_1}.
The primary non-survey methods employed are the location quotient (LQ) and RAS methods.
These methods require less data, facilitating the estimation of input coefficients even in small regions with limited primary statistics.

The location quotient is defined as the ratio of the industry's share of economic activity in the target economy to its share in a reference economy \citep{Isserman1977}.
The LQ method estimates input coefficients for a target region based on the input coefficients of a reference region and the location quotient.
Derivatives include CILQ, RLQ, and FLQ \citep{Flegg2016}.
Recent derivatives, such as FLQ+ and 2D-LQ, have also been developed \citep{PereiraLopez2020, Flegg2021}.
Studies such as \cite{Bonfiglio2008}, \cite{Flegg2012}, and \cite{Lamonica2018} have verified the accuracy of these derivatives.

The RAS method, initially introduced in Stone's research \citep{Bacharach1970}, estimates the input coefficient matrix for a target region by adjusting the initial matrix using the intermediate demands, intermediate inputs, and gross outputs by industry in that region.
\cite{Hewings1977} demonstrated that the RAS method can estimate input coefficients for the target region with high precision.
Similar to the LQ method, numerous derivatives and improvements have been developed for the RAS method
\citep{Junius2003,Lenzen2007,Lemelin2009,Lenzen2014,Temursho2021}.
Furthermore, studies such as \cite{Hiramatsu2016} and \cite{Holy2023} have sought to improve the RAS method, focusing on regional input coefficients.

However, these major non-survey methods require strong assumptions about economic activity, and their validity has been debated.
\cite{Round1983} was one of the first studies on this topic.
\cite{Miller2022_1} outlined the limitations of non-survey methods that have been identified to date.
According to \cite{Riddington2006}, the LQ method assumes that productivity and consumption per employee are identical across all regions and that there is no cross-hauling of products within the same industry.
For the RAS method, \cite{Miller2022_1} noted that differences in production technology between the referenced and target regions must be decomposed into substitution and fabrication effects.
\footnote{Some view the RAS method merely as a calculation to adjust the matrix to satisfy constraints \citep{Miller2022_1}.}

As non-survey methods require less data, the information is also selective, potentially reducing accuracy compared to methods that use more extensive information.

Moreover, both LQ and RAS methods require additional data.
The LQ method requires selecting the reference region and the economic activity when calculating the location quotient.
The RAS method requires selecting the initial input coefficient matrix and specifying intermediate demands, intermediate inputs, and gross outputs by industry in the target region.
These selections influence estimation accuracy.
For instance, \cite{Fukui2025} demonstrated that the accuracy of input coefficient estimation varies depending on the reference region in FLQ and the initial value setting of the input coefficient matrix in RAS.

In recent years, deep learning has been utilized across many fields.
A neural network is a function formed by aligning simple functions, known as activation functions, and connecting them in layers.
A neural network with many layers and structures is called a deep neural network.
Its structure enables deep neural networks to capture nonlinear relationships between variables, making them particularly useful for accurately predicting target variables.
Training deep neural networks with large datasets is referred to as deep learning.
This serves as the foundational framework for current generative artificial intelligence (AI) and is also frequently employed to predict and evaluate asset prices \citep{Ding2020, Srivinay2022, Chen2023}.
In the field of regional economic research, \cite{Khachiyan2022_1} attempted to predict population and income of inhabited areas in the United States from satellite imagery using deep learning.
\cite{Xie2024_1} predicted GDP growth rates across 21 countries using various models, such as multilayer perceptrons, representation transformers, and long short-term memory networks.

Some studies have sought to incorporate neural networks into the estimation of input coefficients.
\cite{Papadas2002} predicted input coefficients for the United Kingdom using a neural network.
\cite{Pakizeh2022} estimated input coefficients for nine regions in Japan using machine learning methods, including neural networks.
Both studies applied a single neural network model to all input coefficients and the model layers were relatively shallow.
Such models may not sufficiently capture the nonlinear relationships between variables.
Consequently, the studies mentioned did not achieve prediction accuracy that clearly surpassed that of conventional methods.
\cite{DePretis2026_1} proposed a method combining generative adversarial networks (GANs) and residual boosting.
This method achieved higher overall accuracy than the RAS method.
However, at this point, it only applied to input coefficients and assumed ideal conditions in which the row and column sums were known, as in the conventional RAS method.

\cite{Fukui2025} developed a method for estimating input coefficients using deep learning.
Training deep neural networks with small datasets may lead to overfitting and reduced prediction accuracy.
This method addressed overfitting through a data augmentation technique known as ``mixup'', as described by \cite{Zhang2018}.
Data augmentation enabled the use of deeper models than those used in previous studies using neural networks.
Compared to conventional methods, the approach of \cite{Fukui2025} requires weaker assumptions about economic activity and achieved superior precision in estimating input coefficients in the input--output table for Japan.
Furthermore, as it does not require additional estimation or selection of other data, the method exhibits stable accuracy.

While several methods have been developed to estimate input coefficients (i.e., the intermediate input section of an input--output table), few studies have developed standardized methods for estimating other sections, such as final demand, gross value added, and gross output.
Simple estimation methods for these sections utilize the ratio of macroeconomic variables between the target and reference regions.
For instance, estimating gross outputs by industry in a region may involve calculating the ratio of workers in each industry relative to the national average, which is then multiplied by national gross outputs by industry to obtain estimates \citep{Kronenberg2009}.
Although straightforward, this method may not accurately reflect actual regional or industry conditions as it does not account for information beyond the calculated ratios.
Alternatively, some methods estimate a table by supplementing the past table with currently available data \citep{Hosoe2014}.
These methods better reflect regional characteristics; however, their feasibility relies on both data availability and the researcher's knowledge and experience.

Given the current state of regional input--output table estimation, this study proposes a deep learning-based method that extends \cite{Fukui2025} approach to include items beyond input coefficients, such as final demand, gross value added, and gross output.
The generation process for each item in an input--output table is modeled through a deep neural network that takes various regional economic data as inputs.
We anticipate that this approach will yield highly accurate and stable estimates for items in the input--output table.

Conventional methods require strict preconditions, including assumptions about the economic structure, additional data, and analyst expertise.
The method proposed in this study has also the advantage of requiring only relatively lenient assumptions.

Section 2 explains the deep learning-based estimation method for regional input--output tables.
Section 3 estimates the 2015 input--output tables for Japan and its selected cities, and verifies the results.
Section 4 discusses the properties and limitations of the proposed method.
Finally, Section 5 summarizes our method.

\section{Data and method}

This section presents the estimation method for the 2015 input--output tables covering Japan as a whole and cities.
There are two reasons to use regional input--output tables of Japan in this analysis: they reflect regional characteristics and are readily available.
Since each local government compiles these tables independently, they accurately depict regional economic characteristics.
Although there is a time lag before publication, tables for all prefectures are made available online free of charge.
In reality, the regional input--output tables for Japan are utilized in research that analyzes the specific conditions of regions within Japan.
\footnote{For example, see \cite{Itakura2024_1}. There are many other studies using regional input--output tables of Japan, but most of them have been published in Japanese.}

The input--output tables for regions in Japan are published approximately every five years with a delay of about five years.
\footnote{
The 2020 table for Japan, for example, were not published until 2024.
For more details, please refer to the website where the tables were published:
\url{https://www.soumu.go.jp/english/dgpp_ss/data/io/io20_00001.html}
}
The 2020 tables are the most recent input--output tables currently available in Japan.
However, the prefectural tables, which serve as the basis for our model estimation, have not yet been released for several regions. Additionally, few tables are available at the city level.
\footnote{In Japan, regional input--output tables are published independently by each local government, and there is no official list providing an overview of the overall publication status.
The Pan Pacific Association of Input--Output Studies (PAPAIOS) has published its own list at the URL below, but it is in Japanese and has a slight time lag:
\url{https://www.gakkai.ne.jp/papaios/iotable-g.html}
}
Therefore, we focused on the 2015 tables, for which more data are available.

Furthermore, since the data used as explanatory variables can be obtained in bulk from the ``Regional Statistics Database,'' data reproducibility and collection efficiency are ensured.
\footnote{The URL for the Regional Statistics Database is as follows: \url{https://www.e-stat.go.jp/en/regional-statistics/ssdsview}}

The data used to train the model were derived from 45 prefectures and 9 cities in Japan, as shown in Table \ref{Tab:Areas}.
These regions were selected to confirm that the improvement in prediction accuracy of the input coefficient matrix described in \cite{Fukui2025}, on which our method is based, can also be achieved when estimating the entire table for similar regions.
However, to further improve accuracy, we included five additional cities compared to \cite{Fukui2025}.

\begin{table}[H]
    \centering
    \caption{Regions included in the training data.} \label{Tab:Areas}
    \small
    \begin{tabular}{l}
        \toprule
        Prefectures \\
        \midrule
        Hokkaido, Aomori, Iwate, Miyagi, Akita, Yamagata, Fukushima, \\
        Ibaraki, Tochigi, Gunma, Saitama, Chiba, Kanagawa, Niigata, \\
        Toyama, Ishikawa, Fukui, Yamanashi, Nagano, Gifu, Shizuoka, \\
        Aichi, Mie, Shiga, Kyoto, Osaka, Hyogo, Nara, Wakayama, \\
        Tottori, Shimane, Okayama, Hiroshima, Yamaguchi, Tokushima, \\
        Kagawa, Ehime, Kochi, Fukuoka, Saga, Nagasaki, Kumamoto, \\
        Oita, Miyazaki, Kagoshima \\
        \midrule
        Cities \\
        \midrule
        Saitama (Saitama Pref.), Chiba (Chiba Pref.), \\
        Yokohama (Kanagawa Pref.), Kawasaki (Kanagawa Pref.), \\
        Sagamihara (Kanagawa Pref.), Shizuoka (Shizuoka Pref.), \\
        Kyoto (Kyoto Pref.), Hiroshima (Hiroshima Pref), \\
        Fukuoka (Fukuoka Pref.) \\
        \bottomrule
    \end{tabular}
\end{table}

The input--output table used in this study was of the competitive import type, with industries reclassified as shown in Table \ref{Tab:Industries}.
Final demand and gross value added were classified as listed in Tables \ref{Tab:FDItems} and \ref{Tab:GVAItems}, respectively.

\begin{table}[H]
    \centering
    \caption{Industry classification.}\label{Tab:Industries}
    \small
    \begin{tabular}{cl}
        \toprule
        Order & Industry \\
        \midrule
        1 & Agriculture, forestry, and fisheries \\
        2 & Mining \\
        3 & Manufacturing \\
        4 & Construction \\
        5 & Electricity, gas, heat supply, water supply, \\
          & and waste disposal business \\
        6 & Commerce \\
        7 & Finance, insurance, and real estate \\
        8 & Transport and postal services \\
        9 & Information and communications \\
        10 & Public administration \\
        11 & Service industries \\
        12 & Unclassified \\
        \bottomrule
    \end{tabular}
\end{table}

\begin{table}[H]
    \centering
    \caption{Sub-sectors within the final demand sector.}\label{Tab:FDItems}
    \small
    \begin{tabular}{cl}
        \toprule
        Order & Sector \\
        \midrule
        1 & Consumption expenditure outside households \\
        2 & Consumption expenditure (private) \\
        3 & Consumption expenditure of general government \\
        4 & Gross regional fixed capital formation \\
        5 & Increase in stocks \\
        6 & Net exports (or total of net exports and net outflow) \\
        \bottomrule
    \end{tabular}
\end{table}

\begin{table}[H]
    \centering
    \caption{Sub-sectors within the gross value added sector.} \label{Tab:GVAItems}
    \small
    \begin{tabular}{cl}
        \toprule
        Order & Sector \\
        \midrule
        1 & Consumption expenditure outside households \\
        2 & Compensation of employees \\
        3 & Operating surplus \\
        4 & Depreciation of fixed capital \\
        5 & Indirect taxes \\
        6 & (less) Current subsidies \\
        \bottomrule
    \end{tabular}
\end{table}

In our method, net exports were the target for estimation, defined as the difference between exports and imports.
As noted in the appendix of \cite{Fukui2025}, in a competitive import type input--output table, the sum of exports from two regions does not equal the exports when those regions are combined. The same applies to imports.
This property does not satisfy the criteria for data augmentation in our method, which will be elaborated upon later.
Thus, exports and imports are unsuitable for estimation by our method.
Conversely, the criteria hold for net exports, making them suitable for our model.

In estimating input--output tables for cities, we focused on the sum of net exports and the net outflow to other regions within the country. Net outflow is defined as the difference between outflow and inflow for the city.

The data used for this study are presented in Table \ref{Tab:OrigData}.
\footnote{The data in Table \ref{Tab:OrigData} regarding the regions of this study (Table \ref{Tab:Areas}) are available on GitHub \citep{Fukui2024_1}.}
In Table \ref{Tab:OrigData}, the subscripts $i$ and $j$ represent the order of the industries in Table \ref{Tab:Industries}.
The subscripts $g$ and $h$ denote the order of final demand in Table \ref{Tab:FDItems} and gross value added in Table \ref{Tab:GVAItems}, respectively.
Furthermore, $k$ indicates regions, and $c$ denotes industry classifications.
In the Economic Census, for minor industry classifications, $c = 1, \ldots, 619$ and for major classifications, $c = 1, \ldots, 17$.
For $\text{CFArea}_{c,k}$ in our analysis, $c$ covers the categories from ``d. Agricultural, forestry and fishery buildings'' to ``r. Buildings that can not be included in other categories.''
\footnote{For details, please refer to Chapter 2, Section 3, ``3) Types of Residential and Industrial Structures'' in the Ministry of Land, Infrastructure, Transport and Tourism's ``Construction Statistics Guidebook.'' This guidebook is available on the following website: \url{https://www.mlit.go.jp/toukeijouhou/chojou/csg/csg_f.htm}.}

\begin{table}[H]
    \begin{threeparttable}
    \centering
    \caption{Data used in this study, grouped by source. The definitions of minor and major classifications follow those of the Economic Census.}\label{Tab:OrigData}
    \footnotesize
    \begin{tabular}{lll}
        \toprule
        Variable & Data & Source \\
        \midrule
        $A_{i,j,k}$ & Intermediate input & 2015 regional input-- \\
         &  (12 industries, million yen) & output tables \\
        $D_{i,g,k}$ & Final demand (12 industries, & published by \\
         & 6 sub-sectors, million yen) & each local government \\
        $V_{i,h,k}$ & Gross value added (12 industries, & \\
         &  6 sub-sectors, million yen) & \\
        $Y_{i,k}$ & Gross output & \\
         &  (12 industries, million yen) & \\
        \midrule
        $\text{Firm}_{c,k}$ & Number of establishments & 2014 Economic Census \\
         & (major classification) & for Business Frame \\
        $\text{SFirm}_{c,k}$ & Number of establishments & \\
         & (minor classification) & \\
        $\text{SEmp}_{c,k}$ & Number of employees & \\
         & (minor classification) & \\
        \midrule
        $\text{MUnitAg}_{k}$ & Number of agriculture & 2015 Census of \\
         & management entities & Agriculture and Forestry \\
        $\text{MUnitFo}_{k}$ & Number of forestal & \\
         & management entities \\
        \midrule
        $\text{ProductsCr}_{k}$ & Production value\tnote{a} & 2015 Statistics of Agri- \\
         & (crop subtotal, 100 million yen) & cultural Income Produced \\
        $\text{ProductsAn}_{k}$ & Production value\tnote{a} & and 2015 Agricultural \\
         & (livestock subtotal, 100 million yen) & Output by Municipality \\
         & & (Estimates) \\
        \midrule
        $\text{CFArea}_{c,k}$ & Total floor area of buildings started & 2015 Building Starts \\
        & (non-residential, $\text{m}^2$) &  \\
        \midrule
        $\text{VA}_{c,k}$ & Value added (major & 2015 Statistical Observa- \\
         & classification, million yen) & tions of Prefectures and \\
        $\text{Sales}_{c,k}$ & Sales (major classification, million yen) & 2015 Statistical Observa- \\
        $\text{Income}_k$ & Taxable income (thousand yen) & tions of Municipalities \\
        $\text{TP}_k$ & Number of taxpayer & \\
        $\text{PopLF}_k$ & Population in labor force & \\
        $\text{Unemp}_k$ & Number of unemployed &  \\
        $\text{Pop15}_k$ & Population aged 15 and over\tnote{b} & \\
        \bottomrule
    \end{tabular}
    \begin{tablenotes}
    \item[] \hspace*{-1em}\textit{Note:}
    \item[a] The Ministry of Agriculture, Forestry and Fisheries has estimated the values of each city.
    \item[b] The values were calculated by the author based on the source data.
    \end{tablenotes}
    \end{threeparttable}
\end{table}

Data augmentation was performed on the data in Table \ref{Tab:OrigData} collected from the regions in Table \ref{Tab:Areas}, to generate training data for the model.
Using these regional data directly in deep learning could lead to overfitting and reduced prediction accuracy due to the small dataset size.
To mitigate this risk of overfitting, \cite{Fukui2025} employed a data augmentation technique, ``mixup'', proposed by \cite{Zhang2018}.
In applying this technique, \cite{Fukui2025} posited the following assumptions:
For each quantitative variable underlying the model variables, the value measured by combining multiple regions equals the sum of the values across those regions, and the value measured by scaling a region equals the scaled value of the region.
These assumptions established prior knowledge that ``for a virtual region obtained by linear interpolation (consisting of the above combination and scaling), the feature vectors should lead to an associated target'' \citep{Fukui2025}.
Using this prior knowledge, we generated data for virtual regions by linearly combining data vectors from multiple regions, thereby increasing the data volume.

During the augmentation process of this study, two to five regions were randomly selected from those in Table \ref{Tab:Areas}.
The weights for the linear combination of these regions were generated randomly from a Dirichlet distribution with all elements of the parameter vector set to one.
This ensures that the sum of the weights is one while enabling uniform sampling \citep{Frigyik2010_1}.
\footnote{Regions in an inclusive relationship, such as Hokkaido and Sapporo City, were not selected concurrently, as interpreting their combination is extremely difficult.
}

Moreover, this study introduced data scaling from \cite{Fukui2025} to address the decline in prediction accuracy owing to the scale difference between the trained and predicted regions.
As most observations were prefectures, the data augmented through the aforementioned method reflected the scale of prefectures.
Consequently, neural network models trained on the generated data exhibited lower accuracy in predicting targets, such as the entire country or municipalities.
To address this, quantitative data for the prefectures were converted to per capita figures for the population aged 15 and over in 2015, and this data was augmented. The augmented quantitative data was then multiplied by the population of each prediction target area to yield training data reflecting the scale of the targets.

To obtain the training data for Japan as a whole, the augmented data was multiplied by the population aged 15 and over for Japan in 2015.
To augment data on cities, we first generated a set of uniform random numbers within the range of the minimum and maximum population aged 15 and over across all cities in Japan in 2015.
We then multiplied these random numbers by each observation in the augmented data to obtain the city training data.

The values of the explanatory and target variables were calculated from the augmented data.
As shown in Table \ref{Tab:Variables}, the explanatory variables were derived from the quantitative variables presented in Table \ref{Tab:OrigData}.
The indices of the explanatory variables are identical to those in Table \ref{Tab:OrigData}.

\begin{table}[H]
    \centering
    \caption{Explanatory variables.}\label{Tab:Variables}
    \small
    \begin{tabular}{ll}
        \toprule
        Name & Definition \\
        \midrule
        Number of establishments (minor classification) & $\text{SFirm}_{c,k}$ \\
        Proportion of establishments by industry & $\text{SFirm}_{c,k} / \sum_c \text{SFirm}_{c,k}$ \\
        (minor classification) & \\
        Number of employees (minor classification) & $\text{SEmp}_{c,k}$ \\
        Proportion of employees by industry & $\text{SEmp}_{c,k} / \sum_c \text{SEmp}_{c,k}$ \\
        (minor classification) & \\
        Number of agriculture management entities & $\text{MUnitAg}_{k}$ \\
        Number of forestal management entities & $\text{MUnitFo}_{k}$ \\
        Production value (crop subtotal) & $\text{ProductsCr}_{k}$ \\
        Production value (livestock subtotal) & $\text{ProductsAn}_{k}$ \\
        Total floor area of buildings started & $\text{CFArea}_{c,k}$ \\
        Proportion of total floor area of buildings & $\text{CFArea}_{c,k} / \sum_c \text{CFArea}_{c,k}$ \\
        started by industry & \\
        Value added (major classification) & $\text{VA}_{c,k}$ \\
        Value added per establishment & $\text{VA}_{c,k} / \text{Firm}_{c,k}$ \\
         (major classification) & \\
        Sales (major classification) & $\text{Sales}_{c,k}$ \\
        Sales per establishment (major classification) & $\text{Sales}_{c,k} / \text{Firm}_{c,k}$ \\
        Taxable income & $\text{Income}_k$ \\
        Taxable income per taxpayer & $\text{Income}_k / \text{TP}_k$ \\
        Population in labor force & $\text{PopLF}_k$ \\
        Labor force participation rate & $\text{PopLF}_k / \text{Pop15}_k$ \\
        Unemployment rate & $\text{Unemp}_k / \text{PopLF}_k$ \\
        \bottomrule
    \end{tabular}
\end{table}

To eliminate the potential impact of regional economic scale, the ratio of each item in the input--output table to the total of gross outputs was used as the target variable.
Economic scale positively correlates with each item in the input--output table and with multiple quantitative explanatory variables, suggesting a similar correlation between input--output items and the explanatory variables.
When this correlation is pronounced, the trained model may clearly reflect it.
To mitigate the influence of this correlation, the following variables were used as the target variables:

\begin{align*}
y_{i,k} &= Y_{i,k} / \sum_i Y_{i,k} \quad \forall k, \\
a_{i,j,k} &= A_{i,j,k} / \sum_i Y_{i,k} \quad \forall i, j, k, \\
d_{i,g,k} &= F_{i,g,k} / \sum_i Y_{i,k} \quad \forall i, g, k, \\
v_{i,h,k} &= V_{i,h,k} / \sum_i Y_{i,k} \quad \forall i, h, k.
\end{align*}
No model was set for the target variables that had a value of zero in all regions in Table \ref{Tab:Areas} and their predicted values were set to zero.

Consequently, estimating total gross output is required separately for the method used in this study.

Before model training, 100,000 samples were generated using the aforementioned data augmentation.
In accordance with \cite{Geron2025_1}, 80\% (80,000 samples) of these were randomly selected as training data, and the remaining 20\% (20,000 samples) served as test data.
We estimated the model parameters using the training data.
Subsequently, predictions were made on the test data, and the model was verified by calculating the prediction error.

\begin{figure}[tb]
    \includegraphics[width=13cm]{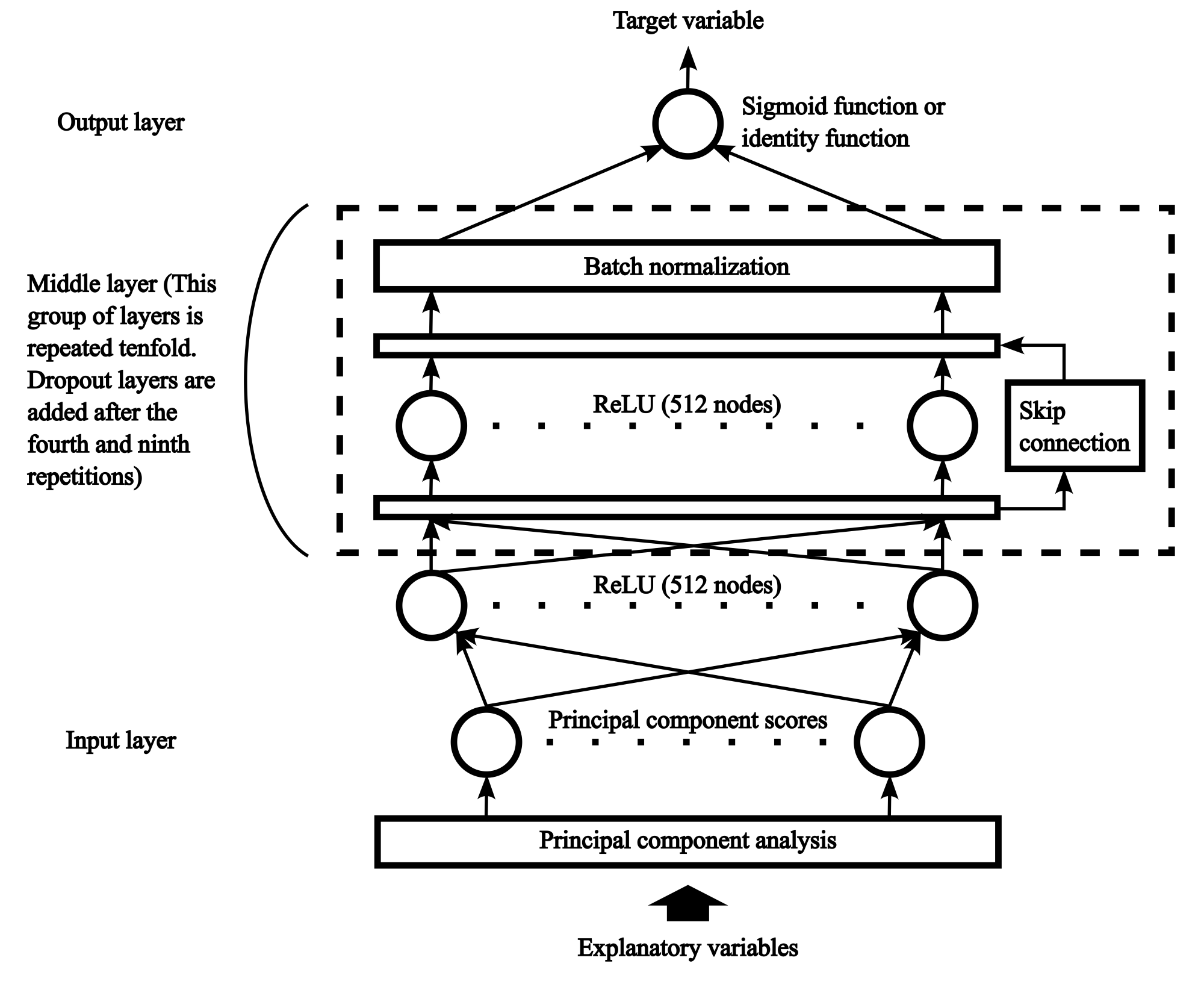}
    \caption{Neural network model employed in this study.}\label{Fig:NN}
\end{figure}

Figure \ref{Fig:NN} illustrates the model for each item in an input--output table.
Before model calculation, the explanatory variables in Table \ref{Tab:Variables} were standardized and converted to principal component scores.
These scores were processed by a rectified linear unit (ReLU) layer with 512 nodes, followed by a stack of ten layers
\footnote{
The following activation function is called the rectified linear unit (ReLU):
\[
y = \max (0, a + \boldsymbol{x}' \boldsymbol{b}),
\]
$\boldsymbol{x}$ is the input vector to ReLU, $y$ is the output from ReLU, $a$ is the bias, and $\boldsymbol{b}$ is the weight for $\boldsymbol{x}$.
}.
Each layer in the stack comprised a ReLU layer (512 nodes), a skip connection, and a batch normalization layer.
The value calculated from the stack was then passed to the output layer to generate the target variable value.

In a batch normalization layer, inputs are standardized with each training data batch, followed by a linear transformation \citep{Ioffe2015}.
In a skip connection, the intermediate layer accounts for the differences between inputs and outputs \citep{He2016}.
Both batch normalization and skip connections enhance the estimation accuracy of model parameters.
Additionally, dropout layers were placed after the fourth and ninth layers in the middle layer to prevent overfitting.
A dropout layer consists of nodes that discard their inputs with a specified probability \citep{Srivastava2014}.

Different output layers were configured based on the target variable.
Assuming the values of each $y_{i,k}$ and $a_{i,j,k}$ are non-negative, a sigmoid function was employed in the output layer.
\footnote{The sigmoid function is equivalent to the logistic function in logistic regression.}
However, when $y_{i,k}$ and $a_{i,j,k}$ values were too small, optimization processes were vulnerable to gradient vanishing in the output layer.
To mitigate this issue, we adopted a target value transformation similar to that described in \cite{Fukui2025}.
The following transformed variable $y^*$ for each of the variables $y_{i,k}$ and $a_{i,j,k}$ was designated as the target variables:
\begin{align}
    y^* &= \frac{y - y_l}{y_u - y_l} \label{eq:yscaling}, \\
    y_l &= \max(0, \min(y) - 0.1 * (\max(y) - \min(y))), \nonumber \\
    y_u &= \min(1, \max(y) + 0.1 * (\max(y) - \min(y))). \nonumber
\end{align}
In this equation, $y$ represents each variable of the $y_{i,k}$ and $a_{i,j,k}$.
Moreover, $\min(y)$ and $\max(y)$ denote the minimum and maximum values of $y$, respectively.
Given that each of $d_{i,g,k}$ and $v_{i,h,k}$ can be both positive and negative, an identity function was implemented as the output layer.
The target variable in the output layer was set to the standardized variable of each $d_{i,g,k}$ and $v_{i,h,k}$.

These transformations for the model variables were applied to the training data.
The values for the transformations (means, standard deviations, minimum and maximum values, principal component loadings, etc.) were utilized in the subsequent prediction phase.

Training was conducted in the following settings:\footnote{For more information on these settings, refer to \cite{Geron2025_1}.}
The mean squared error was configured as the loss function.
The optimization method was stochastic gradient descent with a mini-batch size of 32.
Additionally, Nesterov's accelerated gradient method was applied for parameter updates in stochastic gradient descent, with a parameter of $0.9$.
The learning rate was updated exponentially and periodically, as proposed by \cite{Smith2015}.
The initial value and lower bound were $10^{-6}$, the upper bound was $0.01$, and the step size between the upper and lower bounds was 10.
\footnote{The learning rate corresponds to the step size in the quasi-Newton method.}
Updates to the learning rate and parameter values were performed simultaneously.
The maximum number of epochs for training was set to 200, with early stopping.
First, 20\% of the training data was randomly selected for validation during model training.
The training process terminated when the validation error increased for 10 consecutive epochs.
The parameter values immediately preceding the increase in validation error were adopted as the training result.

Subsequently, the trained neural network was used to predict each element of the regional input--output table.
For the region under prediction, the explanatory variables were standardized, and principal component scores were calculated.
This calculation used the mean, standard deviation, and principal component loadings for each explanatory variable, which were computed using training data.
The principal component scores were then fed into the trained neural network, which produced the transformed elements of the regional table.
As mentioned, since the outputs were also transformed, an inverse transformation was performed using the means, standard deviations, and the values of $y_l$ and $y_u$ of the target variables employed for the training data transformation.
For each variable $y_{i,k}$ and $a_{i,j,k}$, the predicted value $y$ was obtained by performing the inverse transformation of Eq. (\ref{eq:yscaling}) on the output $y^*$ from the model as follows, using the values of $y_l$ and $y_u$ of the training data:
\[
y = y_l + y^* (y_u - y_l).
\]
For each $d_{i,g,k}$ and $v_{i,h,k}$, the inverse transformation from the model output $y^*$ to the predicted value $y$ was as follows:
\[
y = \mu_y + s_y y^*.
\]
Here, $\mu_y$ and $s_y$ are the mean and standard deviation of the training data, respectively.

To estimate the regional input--output table using our trained model, the model's outputs must be multiplied by $\sum_i Y_{i,k}$.

The gross outputs ($Y_{i,k}$) were predicted for each industry individually; thus, the sum of these predicted outputs may not align with the actual total of gross outputs.
To ensure this alignment, each predicted $Y_{i,k}$ was processed as follows.
\[
Y^*_{i,k} = \frac{\tilde{Y}_{i,k}}{\sum_i \tilde{Y}_{i,k}} \sum_i Y_{i,k}
\]
where $\tilde{Y}_{i,k}$ is the gross output for industry $i$ from the model for region $k$.
The processed value $Y^*_{i,k}$ was then used as the predicted gross output.

A matrix balancing procedure was implemented to ensure that the predicted values met the input--output table constraints.
The RAS method, a non-survey technique, is commonly used for matrix balancing.
However, it is not directly applicable to matrices containing negative values.
To address this, \cite{Junius2003} proposed the generalized RAS (GRAS) method, which is based on entropy optimization.
\footnote{\cite{McDougall1999} noted that in situations where the RAS method is applicable, it is equivalent to the entropy optimization method.}
In this study, predicted values were balanced by a entropy optimization based on GRAS under the following two conditions.
First, the row and column sums of the input--output table were equal to the gross outputs by industry.
Second, the total of consumption expenditures outside households in the gross value added sector was equaled the total in the final demand sector.
\footnote{The appendix provides details regarding matrix balancing in this study.}

\section{Results}
Under the configurations described in the previous section, we estimated the 2015 input--output table for Japan, and verified the accuracy of our method.
\footnote{Various computations related to deep learning were performed using Python and the PyTorch library. Data augmentation was implemented in F\#.}
The publicly available input--output tables for Japan are derived from a survey method and exhibit high accuracy, making them suitable benchmarks.

The total gross output used in the estimation process was based on the actual value from the 2015 input--output table.
The model input consisted of the top 60 principal component scores with the highest contribution.

No large differences in prediction errors were observed between the training and test data for any of the trained models; thus, all of the models were deemed valid.

Table \ref{Tab:ErrorsJP} presents indicators of the prediction error for the input--output table of Japan estimated using the method of this study.
These indicators were selected based on \cite{Hosoe2014} and calculated according to the following definitions:
\footnote{In the definition used in \cite{Hosoe2014}, the denominator of the STPE was $\sum X_{i}$. This definition could lead to overestimation when $X_i$ assumes both positive and negative values. Accordingly, this study designated the denominator as $\sum \left| X_i \right|$.}
\begin{align*}
    \text{STPE} &= \sum_i \left| \tilde{X}_{i} - X_{i} \right| / \sum \left| X_i \right|, \\
    \text{MAD} &= \sum_{i} \left| \tilde{X}_{i} - X_{i} \right| / N_1, \\
    \text{U}_2 &= \sqrt{ \sum_{i} \left( \tilde{X}_{i} - X_{i} \right)^2} / \sqrt{\sum_{i} X^2_{i}}, \\
    \text{RMSE} &= \sqrt{\left[ \sum_{i} \left( \tilde{X}_{i} - X_{i} \right)^2 \right] / N_1}, \\
    \text{MAPE} &= (1 / N_2) \sum_{i} \left| (\tilde{X}_{i} - X_{i}) / X_{i} \right|.
\end{align*}
For any indicator, a smaller value indicates a smaller prediction error.
In these definitions, $i$ represents the index for each item in the input--output table, $X_{i}$ is the actual value for $i$, $\tilde{X}_{i}$ is the corresponding estimated value.
$N_1$ denotes the number of items where $X_{i} \neq 0 \ \text{and} \ \tilde{X}_{i} \neq 0$, and $N_2$ is the total number of items whose actual values are neither zero nor empty in the actual input--output table.
\footnote{Items with both actual and estimated values of zero contribute to an underestimation of MAD and RMSE; therefore, these items were excluded from the calculation of these indicators.
Additionally, some items had zero actual values in the national input--output table but not in the prefectural tables.
Our method estimated these items, and their estimates were rarely zero.
Because error rates for these items could not be calculated, they were excluded from the MAPE calculation.}

\begin{table}[H]
    \centering
    \caption{Prediction errors for the 2015 input--output table for Japan.} \label{Tab:ErrorsJP}
    \small
    \begin{tabular}{lccccc}
        \toprule
         & STPE & MAD & $\text{U}_2$ & RMSE & MAPE \\
        \midrule
        All & 0.0536 & 538,816.53 & 0.0446 & 1,406,499.22 & 0.4294 \\
        Intermediate input & 0.0607 & 222,792.53 & 0.0422 & 546,298.93 & 0.1156 \\
        Final demand & 0.0915 & 1,038,292.71 & 0.0845 & 2,032,669.58 & 1.7185 \\
        \ (Excluding net & 0.0649 & 851,980.24 & 0.0675 & 1,788,433.87 & 0.7074 \\
        \ export) & & & & & \\
        \ (Net export only) & 0.4918 & 1,839,436.3 & 0.4168 & 2,853,895.61 & 6.0664 \\
        Gross value added & 0.0477 & 377,953.33 & 0.0414 & 709,621.48 & 0.0904 \\
        Gross output & 0.0312 & 2,642,088.08 & 0.0347 & 4,357,413.12 & 0.0602 \\
        \bottomrule
    \end{tabular}
\end{table}

\begin{figure}[H]
    \centering
    \begin{subfigure}[t]{0.49\linewidth}
        \centering
        \includegraphics[width=6.5cm]{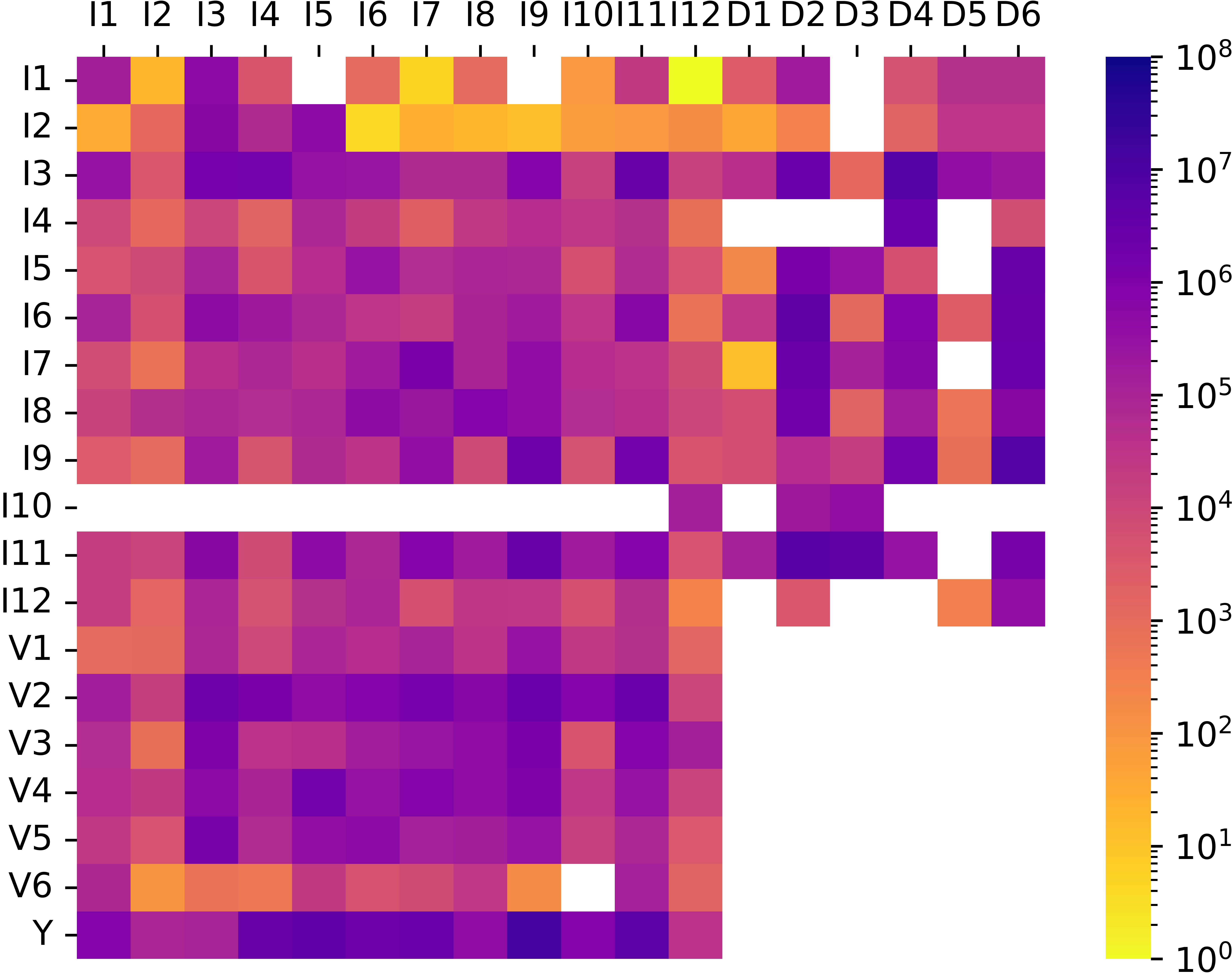}
        \caption{Levels of error. Color-coded on a logarithmic scale.} \label{Fig:HeatMapDiff}
    \end{subfigure}\hfill
    \begin{subfigure}[t]{0.49\linewidth}
        \centering
        \includegraphics[width=6.5cm]{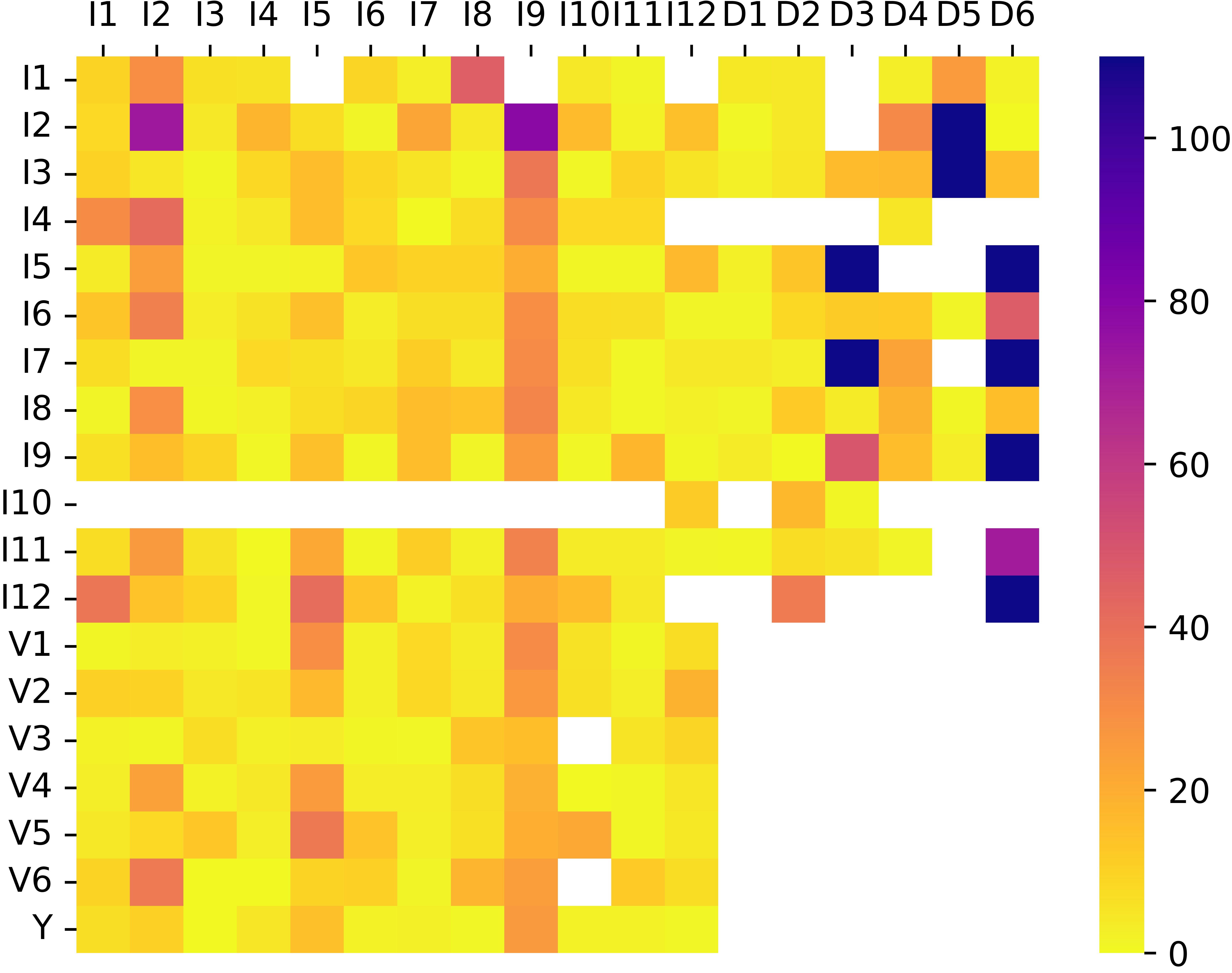}
        \caption{Error rates.} \label{Fig:HeatMapDiffRatio}
    \end{subfigure}
    \caption{Prediction errors (millions of yen) and error rates (\%) for each item in the input--output table for Japan. Each cell represents an absolute value. Items without values are left blank.}
\end{figure}

Figures \ref{Fig:HeatMapDiff} and \ref{Fig:HeatMapDiffRatio} are heatmaps illustrating error levels and error rates for each item in the input--output table for Japan.
These figures follow the structure of the input--output table. Labels ``I1'', $\cdots$, ``I12'' correspond to the order in Table \ref{Tab:Industries}, ``D1'', $\cdots$, ``D6'' correspond to Table \ref{Tab:FDItems}, and ``V1'', $\cdots$, ``V6'' correspond to Table \ref{Tab:GVAItems}. Moreover, the label ``Y'' denotes gross outputs by industry.

Figure \ref{Fig:HeatMapDiff} shows a positive correlation between error magnitude and the gross output of the corresponding industry. Industries with smaller gross output, such as agriculture, forestry, fisheries, and mining, exhibit smaller errors.
In Figure \ref{Fig:HeatMapDiffRatio}, error rates are generally high for final demand items, particularly net exports.

\begin{figure}[H]
    \includegraphics[width=13cm]{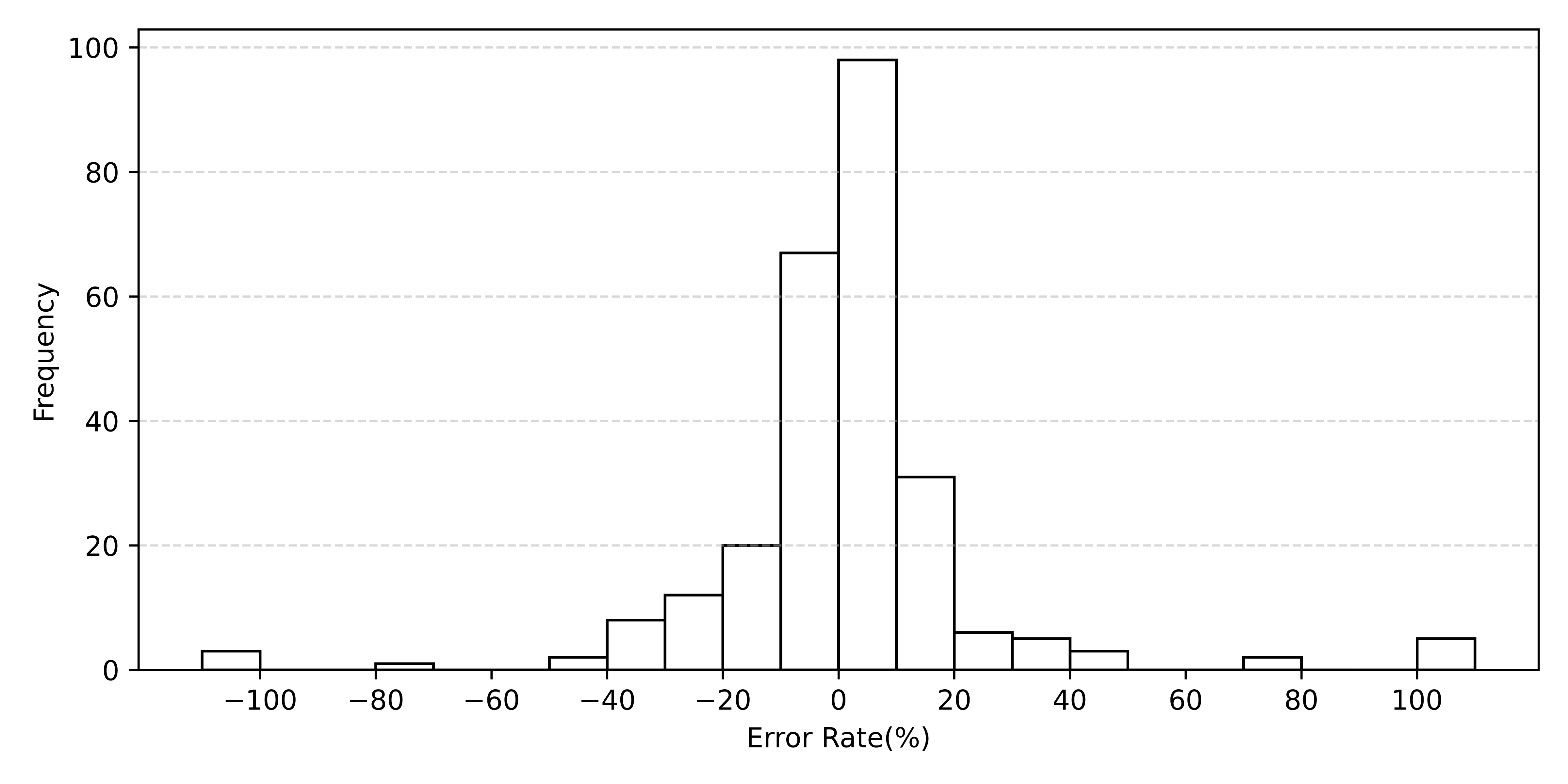}
    \caption{Bar plot of frequencies of prediction error rates for each item in the input--output table for Japan. The rightmost bar represents the frequency with error rates greater than or equal to $100$\%, while the leftmost bar represents the frequency with error rates less than or equal to $-100$\%.} \label{Fig:HistogramRatio}
\end{figure}

Furthermore, Figure \ref{Fig:HistogramRatio} presents a bar plot of error rates.
It shows that 165 items have prediction error rates within the range of [-10\%, 10\%), comprising 55\% of all items (including empty items, for a total of 300).

We compared the accuracy of the proposed method with that of conventional non-survey methods.
As \cite{Morrison1974} stated, the RAS method is generally more accurate than the LQ method when actual row and column totals are available.
However, input--output tables often contain negative values, making direct application of the RAS method inappropriate.
Therefore, we adopted matrix balancing via entropy optimization as a conventional method.
\footnote{This matrix balancing is the same as the one discussed in the previous section. It is explained in detail in the appendix.}
In this matrix balancing, the actual gross outputs by industry in Japan for 2015 served as the row and column totals.
Additionally, the 2011 input--output table for Japan and the 2015 input--output tables for each prefecture were used as the initial values for matrix balancing.

Table \ref{Tab:AccuracyComparison} presents the estimation errors associated with conventional methods.
Using the 2011 table for Japan as the initial value yielded smaller errors than using the 2015 prefecture tables.
Compared to these errors, our method generally produced even smaller errors.

\begin{table}[H]
    \centering
    \caption{Comparison of estimation errors in the input--output table for Japan. The values in the second and third rows were derived using entropy optimization.}\label{Tab:AccuracyComparison}
    \small
    \begin{tabular}{llccccc}
        \toprule
         & STPE & MAD & $\text{U}_2$ & RMSE & MAPE \\
        \midrule
        Our method & 0.0536 & 538,816.53 & 0.0446 & 1,406,499.22 & 0.4294 \\
        Based on 2011 table & 0.0609 & 611,122.95 & 0.05848 & 1,849,283.51 & 0.5673 \\
        Based on 2015 tables & 0.0921 & 936,386.96 & 0.0817 & 2,590,120.89 & 0.5996 \\
        for each prefecture &  &  &  &  &  \\
         (minimum values) &  &  &  &  &  \\
        \bottomrule
    \end{tabular}
\end{table}

Table \ref{Tab:ErrorsJPConventional} presents the estimation errors for each sector using entropy optimization.
The initial value was set to the 2011 input--output table for Japan.
Compared to Table \ref{Tab:ErrorsJP}, the errors for intermediate inputs, final demand (excluding MAPE), and gross value added were larger than those produced by our method.
Our method demonstrated higher prediction accuracy for intermediate inputs, consistent with the results in \cite{Fukui2025}.
However, for net exports in the final demand sector, the errors were smaller than those from our method.

\begin{table}[H]
    \centering
    \caption{Estimation errors of the 2015 input--output table for Japan using entropy optimization with the 2011 table as initial value. Due to the use of actual values for gross outputs by industry, the errors were zero.} \label{Tab:ErrorsJPConventional}
    \small
    \begin{tabular}{lccccc}
        \toprule
         & STPE & MAD & $\text{U}_2$ & RMSE & MAPE \\
        \midrule
        All & 0.0609 & 611,122.95 & 0.0585 & 1,849,283.51 & 0.5673 \\
        Intermediate input & 0.1019 & 382,716.03 & 0.0858 & 1,123,784.12 & 0.3589 \\
        Final demand & 0.1088 & 1,204,079.57 & 0.1307 & 3,146,936.67 & 1.5716 \\
        \ (Excluding net  & 0.0871 & 1,102,510.08 & 0.1222 & 3,245,723.26 & 1.7398 \\
        \ export) & & & & & \\
        \ (Net export only) & 0.4358 & 1,630,671.41 & 0.3932 & 2,692,744.4 & 0.8483 \\
        Gross value added & 0.0862 & 683,274.31 & 0.1017 & 1,744,021.36 & 0.2852 \\
        Gross output & 0.0 & 0.0 & 0.0 & 0.0 & 0.0 \\
        \bottomrule
    \end{tabular}
\end{table}

Overall errors were closer between the two methods compared to sector-specific errors.
However, when taking into account estimating gross outputs by industry, the discrepancy between these methods becomes even more apparent.
In our method, estimation errors occurred for these gross outputs since they were targets of estimation.
In contrast, the conventional method employed actual measured values.
Consequently, as shown in Table \ref{Tab:ErrorsJPConventional}, the errors were zero.
Discrepancies in estimating gross outputs by industry likely contributed to the overall error.
Excluding these outputs, the STPE and MAD were $0.0677$ and $438,261.71$, respectively, using our method, and $0.0991$ and $640,813.13$, respectively, using the conventional method.
Our method also produced smaller values for other indicators.

Using the method developed in this study, we estimated input--output tables for three cities: Sapporo, Gujo, and Okayama.
The estimation utilized the published total gross output for each city, selecting the top 100 principal component scores with the highest contribution as model inputs.
We subsequently compared the input--output tables published by these three cities with our estimates.

\begin{figure}[H]
    \centering
    \begin{subfigure}[t]{0.49\linewidth}
        \centering
        \includegraphics[width=6.5cm]{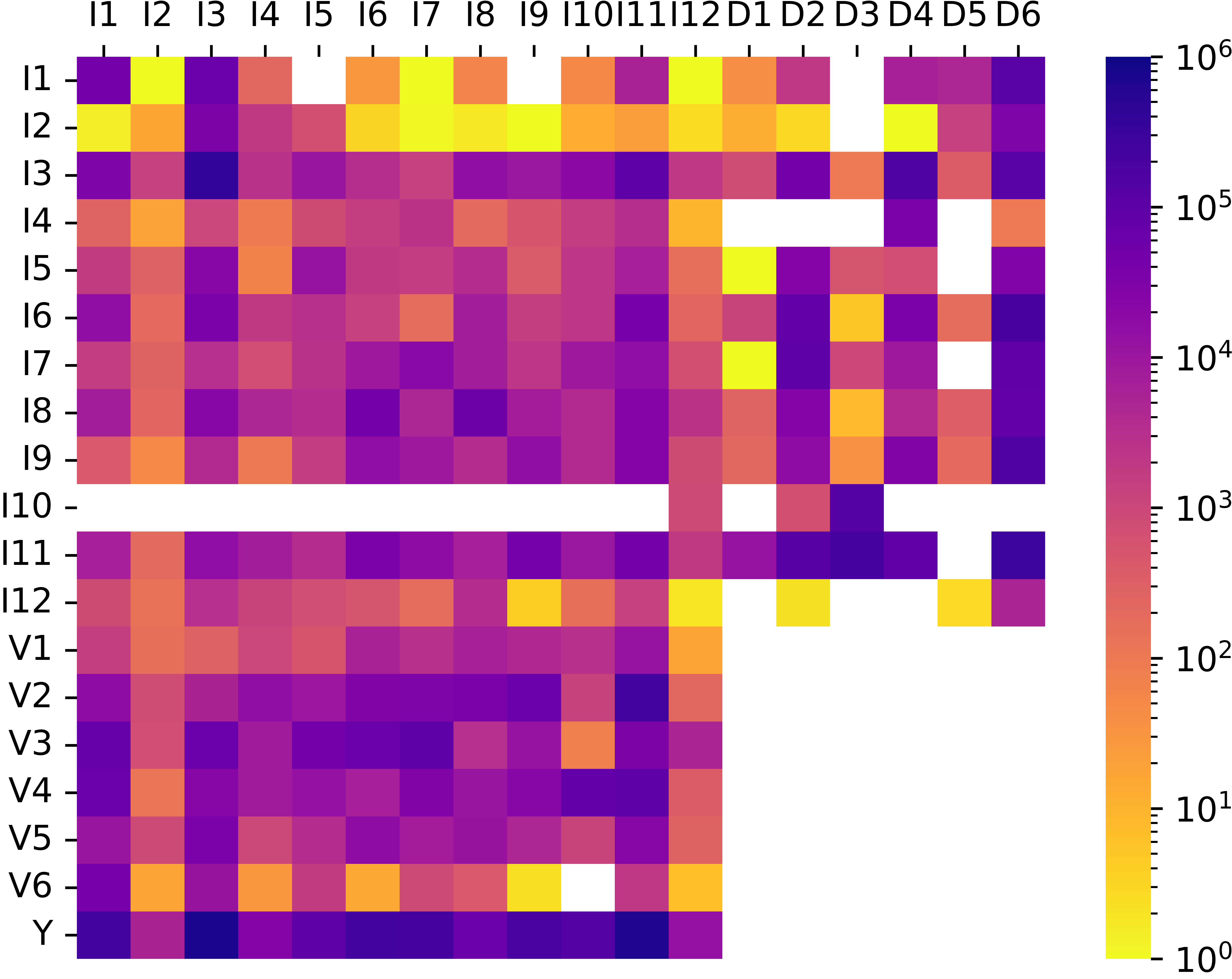}
        \caption{Levels of difference. Color-coded on a logarithmic scale.} \label{Fig:HeatMapDiff01100}
    \end{subfigure}\hfill
    \begin{subfigure}[t]{0.49\linewidth}
        \centering
        \includegraphics[width=6.5cm]{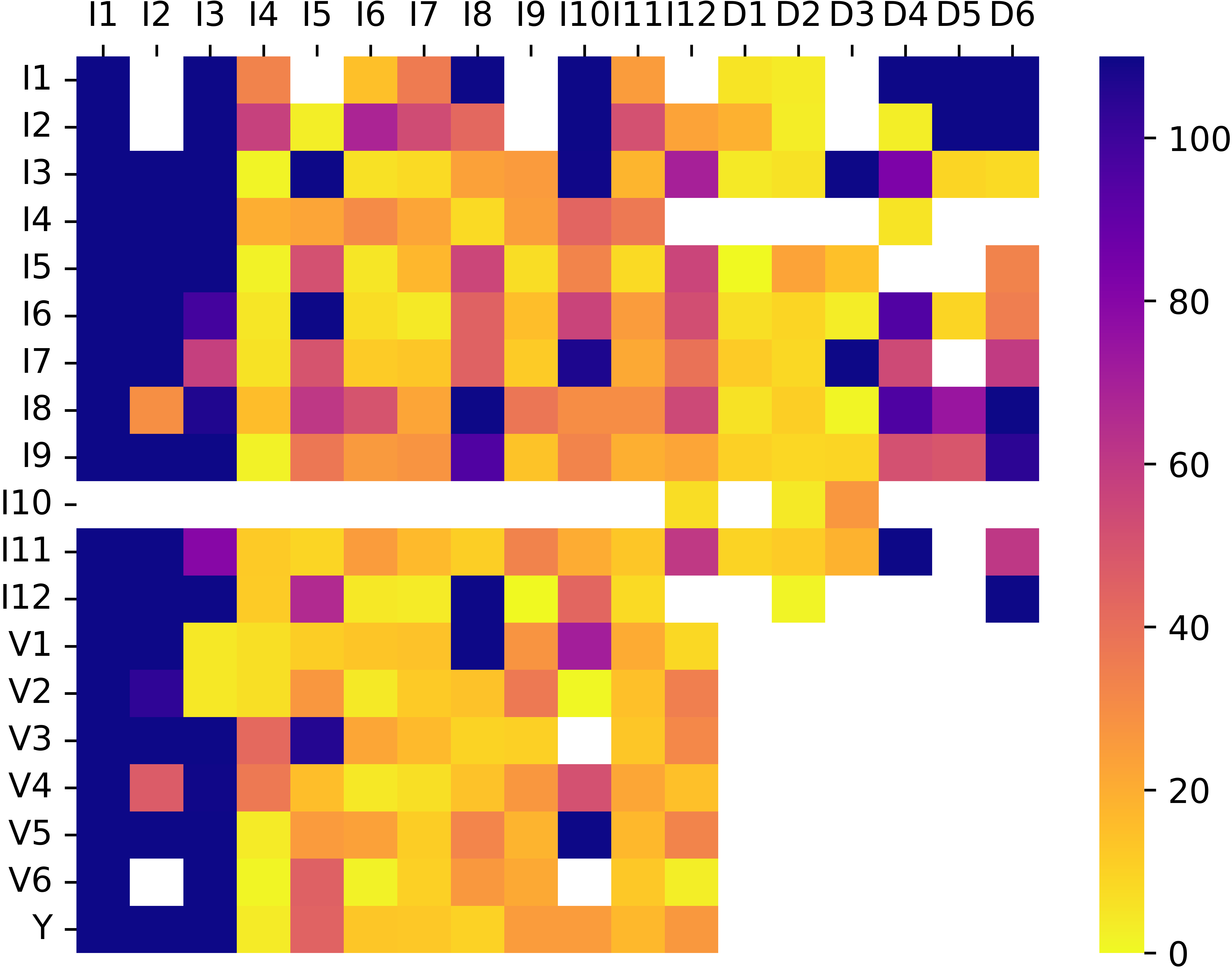}
        \caption{Difference rates.} \label{Fig:HeatMapDiffRatio01100}
    \end{subfigure}
    \caption{Levels of difference (millions of yen) and difference rates (\%) between the estimated and published values for each item in the input--output table for Sapporo City. Each cell represents an absolute value. Items without values are left blank.} \label{Fig:HeatMap01100}
\end{figure}

\begin{figure}[H]
    \centering
    \begin{subfigure}[t]{0.49\linewidth}
        \centering
        \includegraphics[width=6.5cm]{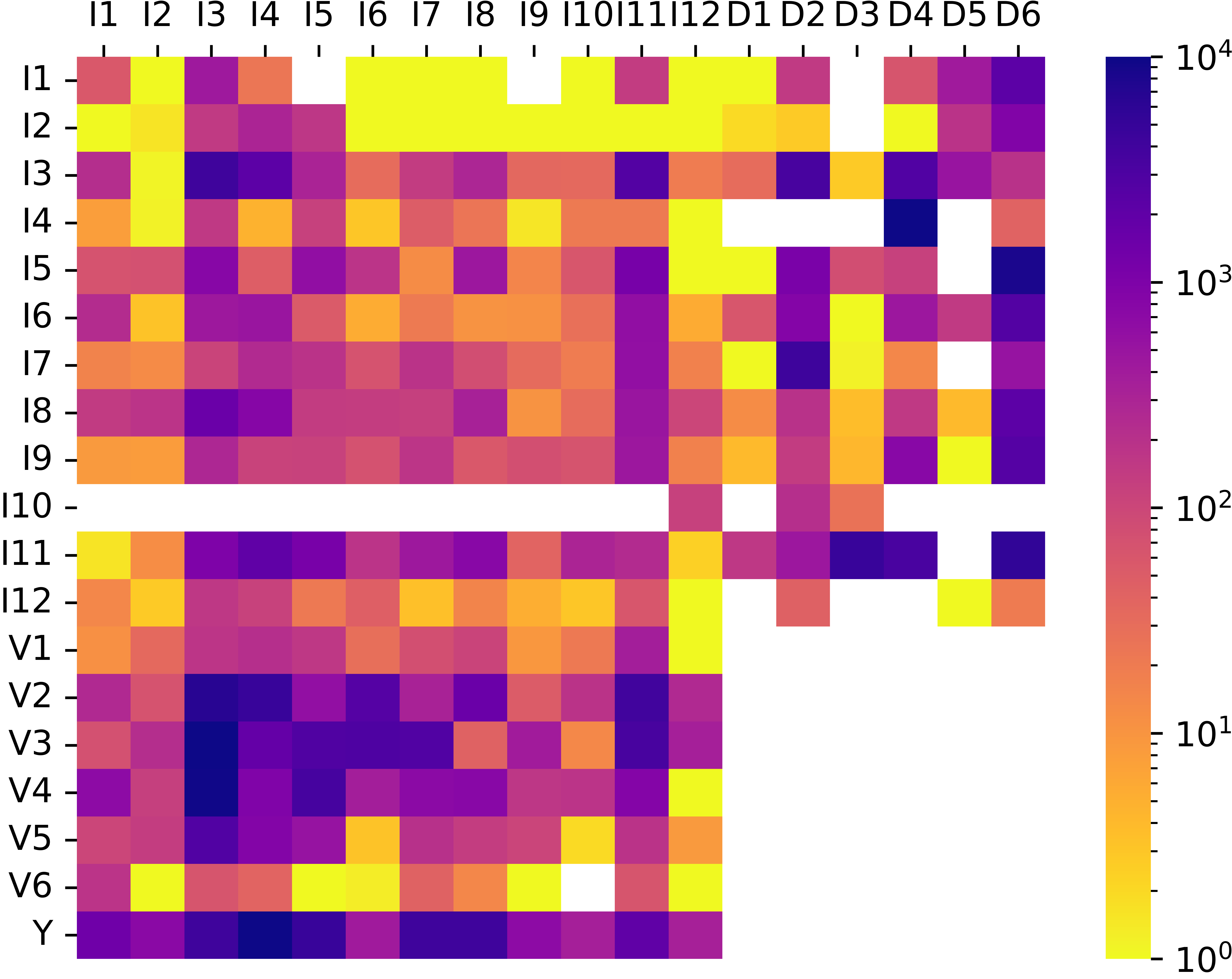}
        \caption{Levels of difference. Color-coded on a logarithmic scale.} \label{Fig:HeatMapDiff21219}
    \end{subfigure}\hfill
    \begin{subfigure}[t]{0.49\linewidth}
        \centering
        \includegraphics[width=6.5cm]{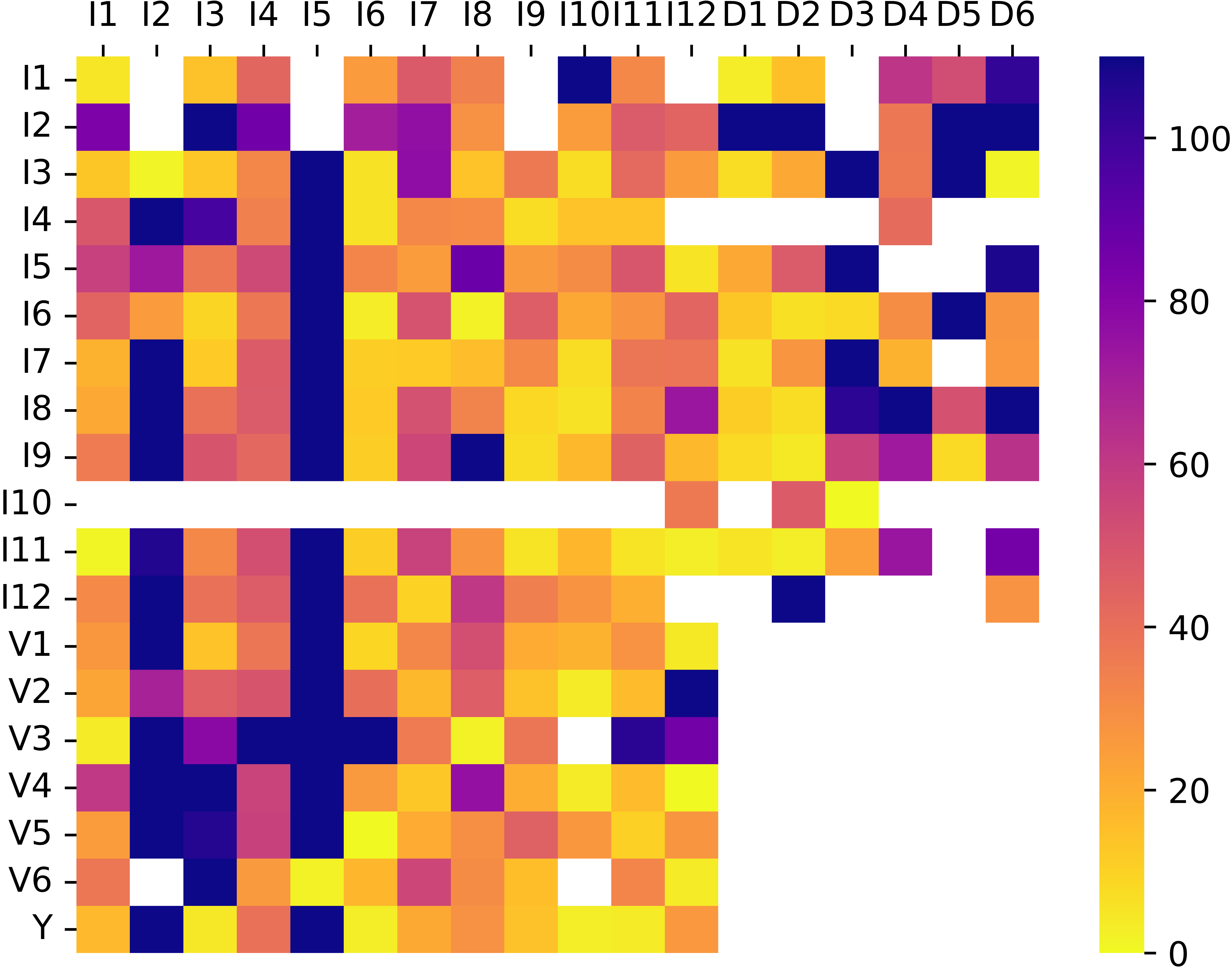}
        \caption{Difference rates.} \label{Fig:HeatMapDiffRatio21219}
    \end{subfigure}
    \caption{Levels of difference (millions of yen) and difference rates (\%) between the estimated and published values for each item in the input--output table for Gujo City. Each cell represents an absolute value. Items without values are left blank.} \label{Fig:HeatMap21219}
\end{figure}

\begin{figure}[H]
    \centering
    \begin{subfigure}[t]{0.49\linewidth}
        \centering
        \includegraphics[width=6.5cm]{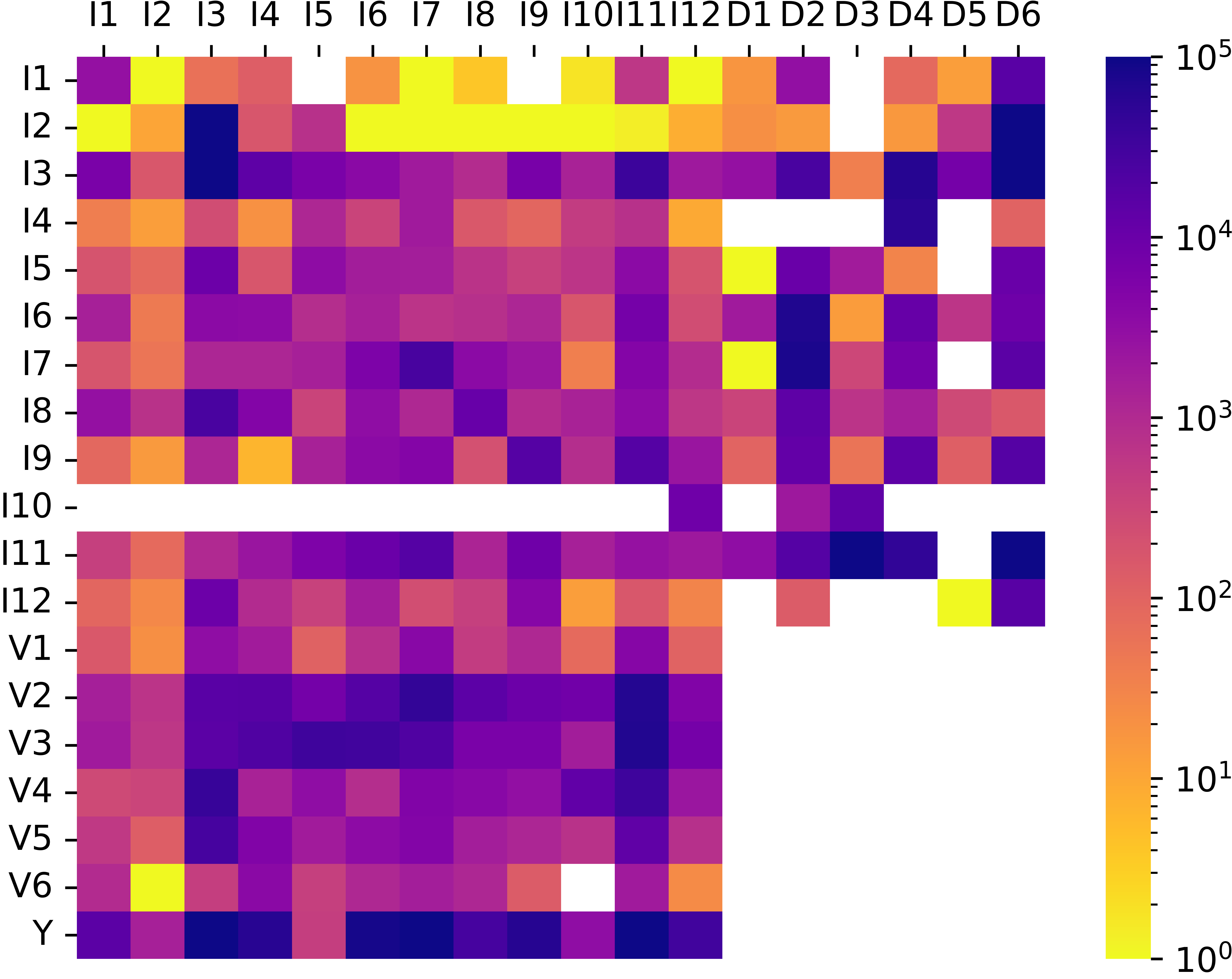}
        \caption{Levels of difference. Color-coded on a logarithmic scale.} \label{Fig:HeatMapDiff33100}
    \end{subfigure}\hfill
    \begin{subfigure}[t]{0.49\linewidth}
        \centering
        \includegraphics[width=6.5cm]{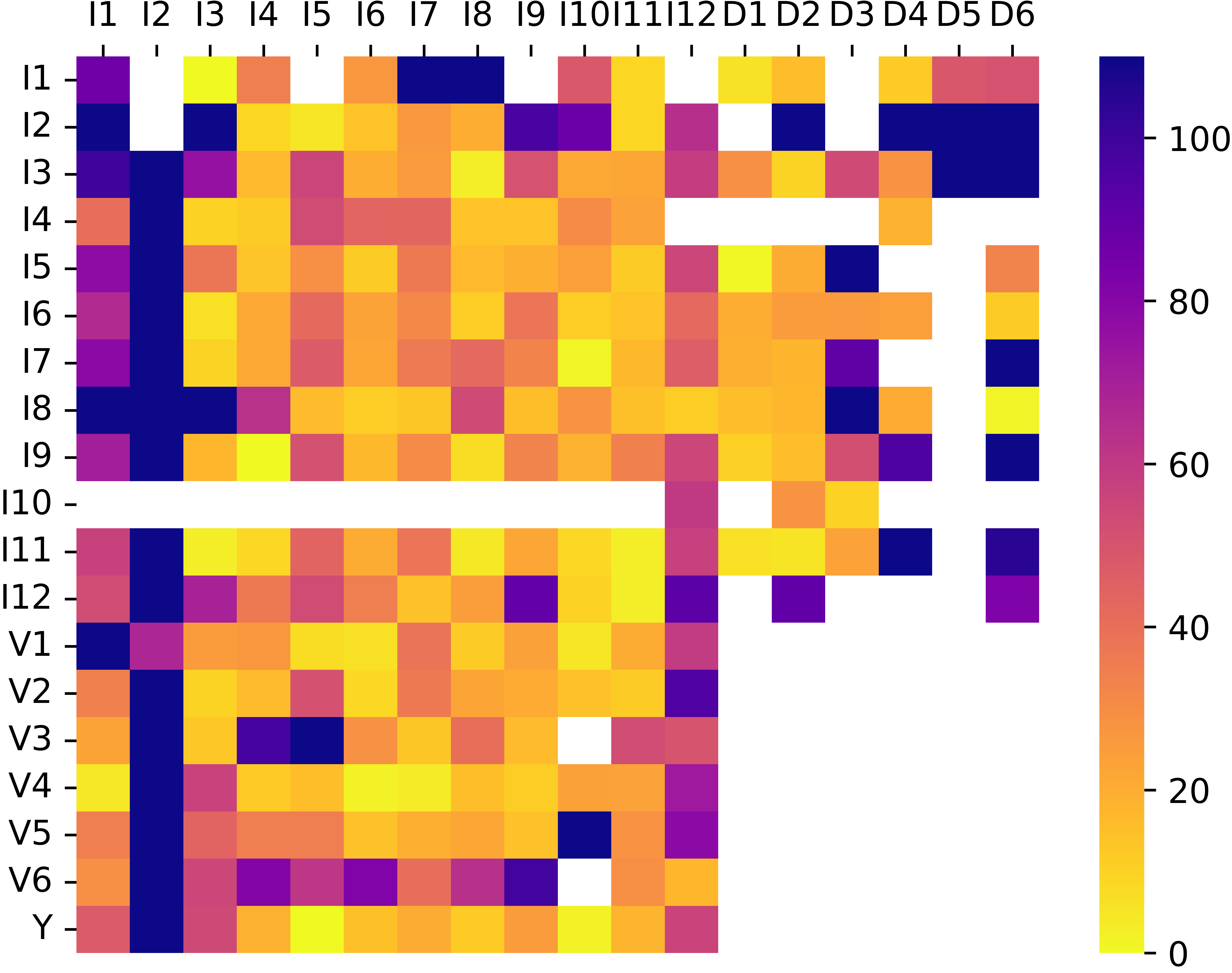}
        \caption{Difference rates.} \label{Fig:HeatMapDiffRatio33100}
    \end{subfigure}
    \caption{Levels of difference (millions of yen) and difference rates (\%) between the estimated and published values for each item in the input--output table for Okayama City. Each cell represents an absolute value. Items without values are left blank.} \label{Fig:HeatMap33100}
\end{figure}

The estimates derived from our method differed relatively substantially from the published values for certain items.
The sub-tables in Table \ref{Tab:DiffsCities} outline the differences between the estimated and published values for the three cities, measured by STPE, MAD, $\text{U}_2$, RMSE, and MAPE.
The sub-figures in Figure \ref{Fig:HeatMap01100}, \ref{Fig:HeatMap21219}, and \ref{Fig:HeatMap33100} present heatmaps of the differences and difference rates for each item in the input--output tables for each city.

\begin{table}[H]
    \centering
    \caption{Differences between the estimated and published values of the input--output tables for the three cities.} \label{Tab:DiffsCities}
    \begin{subtable}{0.99\linewidth}
        \centering
        \caption{Sapporo City}
        \small
        {\begin{tabular}{lccccc}
            \toprule
             & STPE & MAD & $\text{U}_2$ & RMSE & MAPE \\
            \midrule
            All & 0.2424 & 29,799.66 & 0.2233 & 83,608.49 & 3.1939 \\
            Intermediate input & 0.3829 & 11,869.63 & 0.5469 & 40,219.63 & 2.6315 \\
            Final demand & 0.2176 & 42,492.53 & 0.1927 & 78,554.27 & 2.6495 \\
            \ (Excluding net export) & 0.1584 & 26,475.88 & 0.1516 & 55,531.03 & 3.0604 \\
            \ (Net export only) & 0.3523 & 111,364.1 & 0.2528 & 139,446.08 & 0.8829 \\
            Gross value added & 0.2019 & 20,342.04 & 0.1663 & 40,397.84 & 4.8545 \\
            Gross output & 0.2418 & 214,892.03 & 0.2267 & 312,069.42 & 1.9088 \\
            \bottomrule
            \vspace*{1mm}
        \end{tabular}}
    \end{subtable}
    \hfill
    \begin{subtable}{0.99\linewidth}
        \centering
        \caption{Gujo City}
        \small
        {\begin{tabular}{lccccc}
            \toprule
             & STPE & MAD & $\text{U}_2$ & RMSE & MAPE \\
            \midrule
            All & 0.2826 & 776.94 & 0.2305 & 1,964.12 & 1.9202 \\
            Intermediate input & 0.2669 & 249.33 & 0.1855 & 606.68 & 0.8396 \\
            Final demand & 0.3107 & 1,160.12 & 0.344 & 2,510.61 & 3.2751 \\
            \ (Excluding net export) & 0.2414 & 850.79 & 0.2975 & 2,228.24 & 2.0114 \\
            \ (Net export only) & 0.5386 & 2,490.19 & 0.5417 & 3,472.35 & 8.7093 \\
            Gross value added & 0.5217 & 1,063.15 & 0.528 & 2,418 & 3.0375 \\
            Gross output & 0.1361 & 2,890.83 & 0.1271 & 4,204.89 & 0.6769 \\
            \bottomrule
            \vspace*{1mm}
        \end{tabular}}
    \end{subtable}
    \hfill
    \begin{subtable}{0.99\linewidth}
        \centering
        \caption{Okayama City}
        \small
        {\begin{tabular}{lccccc}
            \toprule
             & STPE & MAD & $\text{U}_2$ & RMSE & MAPE \\
            \midrule
            All & 0.3138& 15,749.46 & 0.3793 & 56,149.35 & 14.6972 \\
            Intermediate input & 0.4221 & 6,546.51 & 0.7138 & 30,807.35 & 1.031 \\
            Final demand & 0.4101 & 28,492.73 & 0.5306 & 72,969.88 & 73.309 \\
            \ (Excluding net export) & 0.2022 & 14,911.8 & 0.1996 & 29,825.28 & 92.1305 \\
            \ (Net export only) & 1.5653 & 80,100.25 & 1.9597 & 148,922.18 & 1.7873 \\
            Gross value added & 0.2188 & 8,855.14 & 0.1888 & 16,832.63 & 1.7445 \\
            Gross output & 0.2579 & 102,391.1 & 0.3246 & 185,810.67 & 0.4417 \\
            \bottomrule
        \end{tabular}}
    \end{subtable}
\end{table}

The STPE, $\text{U}_2$, and MAPE in Table \ref{Tab:DiffsCities}, which are independent of the magnitude of amounts, shows that their values for the three cities were generally higher than those for Japan.
Notably, Okayama City's final demand exhibited a large MAPE.
Figures \ref{Fig:HeatMap01100}, \ref{Fig:HeatMap21219}, and \ref{Fig:HeatMap33100}, reveal positive correlations between gross outputs by industry and the magnitudes of these differences, similar to the estimation results for Japan's input--output table.
Furthermore, these figures indicate a tendency toward higher difference rates of input in mining across all cities.
Sapporo City demonstrated high difference rates of input in agriculture, forestry, and fisheries, as well as in manufacturing.
Gujo City showed high difference rates of input in electricity, gas, and water supply.

However, it is important to note that the above results do not necessarily reflect estimation accuracy.
Therefore, a definitive conclusion about the validity of the method developed in this study for estimating input--output tables for three cities in Japan could not be reached.
Estimation accuracy is determined by comparing estimates with true values or values close to them.
The input--output tables published by these cities were estimated using non-survey or hybrid methods because insufficient primary statistics were available for their compilation.
These methods may exhibit larger estimation errors than the survey method and do not guarantee proximity to the true values.
Consequently, the estimation accuracy of the input--output tables for these cities could not be ascertained by comparing the estimates with the published values.

\section{Discussion}
The proposed method achieved higher accuracy in estimating the input--output table for Japan compared to conventional methods.
When estimating the 2015 input--output table for Japan, the conventional matrix balancing relied on the 2011 table and 2015 gross outputs by industry, both obtained through survey methods, and thus represented an estimation under ideal assumptions.
In comparison, our method generally produced lower errors than the conventional method based on these ideal assumptions.
This is a noteworthy observation.

The high accuracy of this method is primarily due to the implementation of deep learning.
This study addressed overfitting by generating virtual regions through data augmentation.
When the model was trained on the augmented data, the prediction accuracy for the input--output table for Japan was satisfactory, indicating that data augmentation effectively mitigated overfitting.

Compared to a more precise method, this study achieved equivalent accuracy.
\cite{Hosoe2014} estimated the 2005 input--output table for Japan by partially revising a past table with the available 2005 data and then balancing it.
The results showed a MAPE of 44\%, and approximately 53\% of the 486 items had an error rate within $[-10\%, 10\%)$.
By contrast, our method yielded a MAPE of 43\% and an error rate within $[-10\%, 10\%)$ for 55\% of the 300 items.
Based solely on these metrics, the prediction accuracy of the two methods can be considered comparable.

The method proposed in this study circumvents the instability associated with selecting reference values, which is necessary for conventional estimation methods.
Both entropy optimization and the RAS method require an initial table for calculation, and the LQ method requires selecting a reference region and variables to derive location quotients.
These selections can affect estimation accuracy.
As shown in Table \ref{Tab:AccuracyComparison}, conventional entropy optimization exhibited varying estimation accuracy based on the initial table selected.
By contrast, while our method requires the establishment of explanatory variables, it does not require selecting reference values as conventional methods do.
Therefore, it does not exhibit the accuracy fluctuations observed in conventional methods.

Compared with entropy optimization and the RAS method, our approach requires fewer specified values in advance.
For instance, the entropy optimization requires gross outputs by industry to serve as row and column sums during balancing.
By contrast, our method merely requires the total of gross outputs, which is generally less challenging to estimate than gross outputs for each industry.
For Japanese municipalities, estimating total gross output is relatively straightforward due to the availability of various data sources, such as the Economic Census.

Although our method demonstrated higher accuracy than conventional approaches in estimating the input--output table for Japan, its efficacy in smaller regions, such as cities, remains unclear.
In three Japanese cities, we observed relatively large differences between our estimates and the published values for certain items.
The regions used for training were predominantly prefectures, which may limit the model's ability to reflect city properties accurately.
However, it is important to note that the published values for these cities are also estimates and may not accurately reflect true values.

The accuracy of certain final demand items, particularly net exports, could be improved.
One potential enhancement would be to introduce explanatory variables that correlate more closely with these items.
However, this study could not identify suitable variables for small-area estimation.

This study's method does not completely supplant conventional methods, as it cannot satisfy the constraints of input--output tables on its own; the matrix balancing methods must be used alongside it.
Furthermore, our method has limitations in directly estimating exports and imports, necessitating alternative methods for their estimation, such as the one proposed by \cite{Fujimoto2019_1}.

\section{Conclusion}
This study proposes a method for estimating regional input--output tables using deep learning.
This method achieved higher estimation accuracy than the matrix balancing method, which assumes that the row sums and column sums of the input--output tables are known.

Many regions and countries make it difficult to conduct sufficient surveys to compile input--output tables.
Although our method has limitations and room for improvement, it can provide a basis for creating more detailed tables in these regions.

Once the method is established and regional input--output tables can be estimated with high accuracy, quantitative analyses based on the tables, such as evaluating economic impacts and performing applied general equilibrium analyses, can be extended to many regions.
As a result, it will be possible to accurately assess regional conditions and conduct detailed policy deliberations.

However, our method has a low estimation accuracy for final demand items, which requires improvement.
We plan to address this issue by improving our model.

Furthermore, to address the delay in publishing input--output tables, we will explore ways to apply this method to estimating preliminary figures.

\appendix

\section{Overview of matrix balancing using entropy optimization}

This section provides an overview of the matrix balancing employed in this study.
\footnote{
This overview is based on a study (in Japanese) published by the author in 2025.
}

To ensure that the estimated input--output table satisfies the relevant constraints, matrix balancing was implemented.
The method was based on the entropy optimization formulated in \cite{Chlebus2023}.
The objective function of the GRAS method, as described in \cite{Junius2003} and \cite{Lenzen2007}, was incorporated into the formulation, along with additional constraints regarding non-household consumption expenditure.

Let $K$ be the number of industries in an input--output table, $L$ be the number of sub-sectors in gross value added, and $M$ be the number of sub-sectors in final demand.
The row and column indices of the input--output table are denoted by $i$ and $j$, respectively.
For each item in the table, we denote the value divided by the total of gross outputs as $p_{i,j} \ (1 \le i \le K+L, 1 \le j \le K+M)$, which is estimated using matrix balancing.
An initial input--output table is required for this estimation.
For the initial table, we also use the value divided by the total of gross outputs for each item, denoted by $p^0_{i,j} \ (1 \le i \le K+L, 1 \le j \le K+M)$.

The purpose of matrix balancing is to identify the values of $p_{i,j}$ that minimize the following objective function:
\begin{equation}
    - \sum_{i=1}^{K+L} \sum_{j=1, p^0 \ge 0}^{K+M} p_{i,j} \log \frac{p_{i,j}}{e \cdot p^0_{i,j}} + \sum_{i=1}^{K+L} \sum_{j=1, p^0 < 0}^{K+M} p_{i,j} \log \frac{p_{i,j}}{e \cdot p^0_{i,j}}. \label{eq:objfunc}
\end{equation}
Notation $p^0 \ge 0$ indicates that the sum comprises exclusively of terms where $p^0$ is greater than or equal to zero. Conversely, $p^0 < 0$ denotes that the sum consists only of terms where $p^0_{i,j}$ is less than zero.
Elements in which $i > K$ and $j > K$ are excluded from the objective function due to their absence.

In the minimization process, certain constraints must be satisfied.
In input--output tables, the total input and total demand for an industry are equal to the gross output of that industry. Therefore, the following constraints must be considered:
\begin{align}
    & \sum_{j=1}^{K + M} p_{i,j} = y_i, \quad 1 \le i \le K \label{eq:rowsums} \\
    & \sum_{i=1}^{K + L} p_{i,j} = y_j, \quad 1 \le j \le K \label{eq:colsums},
\end{align}
where $y_i$ is the gross output of industry $i$ divided by the total of the gross outputs. When $i = j$, it follows that $y_i = y_j$.

In the input--output tables of Japan, it is customary to include a sub-sector of consumption expenditure outside households in both final demand and gross value added.
The totals of consumption expenditure outside households across all industries in their two sectors must be equal.
Therefore, the following constraint is imperative:
\begin{equation}
    \sum_{j=1}^K p_{\phi, j} = \sum_{i=1}^K p_{i, \phi}, \label{eq:nonhousehold}
\end{equation}
where $\phi \ (\phi > K)$ represents the index of the consumption expenditure outside households sector in final demand and gross value added.

From the objective function and constraints enumerated above, the first-order conditions for optimization with Lagrange multipliers specify that the estimates of $p_{i,j}$, obtained through matrix balancing, are as follows:
\begin{align}
    & \text{For intermediate input} \ (1 \le i \le K, 1 \le j \le K) \text{:} \nonumber \\
    & \qquad p_{i,j} = \begin{cases}
        p^0_{i,j} \exp (\lambda_i + \tau_j), \quad p^0_{i,j} \ge 0 \\
        p^0_{i,j} \exp (-\lambda_i - \tau_j), \quad p^0_{i,j} < 0 \\
    \end{cases} \label{eq:p_a} \\
    & \text{For final demand without consumption expenditure outside households} \nonumber \\
    & \ (1 \le i \le K, K < j \le K + M, j \neq \phi) \text{:}\nonumber \\
    & \qquad p_{i,j} = \begin{cases}
        p^0_{i,j} \exp (\lambda_i), \quad p^0_{i,j} \ge 0 \\
        p^0_{i,j} \exp (-\lambda_i), \quad p^0_{i,j} < 0 \\
    \end{cases} \label{eq:p_f} \\
    & \text{For gross value added without consumption expenditure outside} \nonumber \\
    & \text{households} \ (K < i \le K+L, i \neq \phi, 1 \le j \le K) \text{:} \nonumber \\
    & \qquad p_{i,j} = \begin{cases}
        p^0_{i,j} \exp (\tau_j), \quad p^0_{i,j} \ge 0 \\
        p^0_{i,j} \exp (-\tau_j), \quad p^0_{i,j} < 0 \\
    \end{cases} \label{eq:p_v} \\
    & \text{For consumption expenditure outside households in final demand} \nonumber \\
    & (1 \le i \le K, j = \phi) \text{:} \nonumber \\
    & \qquad p_{i,\phi} = \begin{cases}
        p^0_{i,\phi} \exp (\lambda_i - \eta), \quad p^0_{i,\phi} \ge 0 \\
        p^0_{i,\phi} \exp (-\lambda_i + \eta), \quad p^0_{i,\phi} < 0 \\
    \end{cases} \label{eq:p_phic} \\
    & \text{For consumption expenditure outside households in gross value added} \nonumber \\
    & (i = \phi, 1 \le j \le K) \text{:} \nonumber \\
    & \qquad p_{\phi,j} = \begin{cases}
        p^0_{\phi,j} \exp (\tau_j + \eta), \quad p^0_{\phi,j} \ge 0 \\
        p^0_{\phi,j} \exp (-\tau_j - \eta), \quad p^0_{\phi,j} < 0, \\
    \end{cases} \label{eq:p_phir}
\end{align}
where $\lambda_i$, $\tau_j$, and $\eta$ are the Lagrange multipliers associated with the constraint equations of Eq. (\ref{eq:rowsums}), (\ref{eq:colsums}), and (\ref{eq:nonhousehold}), respectively.

The expressions for $\exp(\lambda_i)$, $\exp(\tau_j)$, and $\exp(\eta)$ are derived from the above equations, along with Eq. (\ref{eq:rowsums}), (\ref{eq:colsums}), and (\ref{eq:nonhousehold}):
\begin{align}
    & \left( \sum_{j = 1, p^0 \ge 0}^K p^0_{i,j} \exp(\lambda_i) \exp(\tau_j) + \sum_{j = 1, p^0 < 0}^K \frac{p^0_{i,j}}{\exp(\lambda_i) \exp(\tau_j)} \right) \nonumber \\
    & \quad + p^0_{i,\phi} \frac{\exp(\lambda_i)}{\exp(\eta)} + \left( \sum_{\substack{j = K+1, j \neq \phi, \\ p^0 \ge 0}}^{K + M} p^0_{i,j} \exp(\lambda_i) + \sum_{\substack{j = K+1, j \neq \phi, \\ p^0 < 0}}^{K + M} \frac{p^0_{i,j}}{\exp(\lambda_i)} \right) \nonumber \\
    & = y_i, \ (1 \le i \le K, \ p^0_{i,\phi} \ge 0) \label{eq:lambda1_p} \\
    & \left( \sum_{j = 1, p^0 \ge 0}^K p^0_{i,j} \exp(\lambda_i) \exp(\tau_j) + \sum_{j = 1, p^0 < 0}^K \frac{p^0_{i,j}}{\exp(\lambda_i) \exp(\tau_j)} \right) \nonumber \\
    & \quad + p^0_{i,\phi} \frac{\exp(\eta)}{\exp(\lambda_i)} + \left( \sum_{\substack{j = K+1, j \neq \phi, \\ p^0 \ge 0}}^{K + M} p^0_{i,j} \exp(\lambda_i) + \sum_{\substack{j = K+1, j \neq \phi, \\ p^0 < 0}}^{K + M} \frac{p^0_{i,j}}{\exp(\lambda_i)} \right) \nonumber \\
    & = y_i, \ (1 \le i \le K, \ p^0_{i,\phi} < 0) \label{eq:lambda1_n} \\
    & \left( \sum_{i = 1, p^0 \ge 0}^K p^0_{i,j} \exp(\lambda_i) \exp(\tau_j) + \sum_{i = 1, p^0 < 0}^K \frac{p^0_{i,j}}{\exp(\lambda_i) \exp(\tau_j)} \right) \nonumber \\
    & \quad + p^0_{\phi,j} \exp(\tau_j) \exp(\eta) \nonumber \\
    & \quad + \left( \sum_{\substack{i = K+1, i \neq \phi, \\ p^0 \ge 0}}^{K + L} p^0_{i,j} \exp(\tau_j) + \sum_{\substack{i = K+1, i \neq \phi, \\ p^0 < 0}}^{K + L} \frac{p^0_{i,j}}{\exp(\tau_j)} \right) \nonumber \\
    & = y_j, \ (1 \le j \le K, \ p^0_{\phi,j} \ge 0) \label{eq:lambda2_p} \\
    & \left( \sum_{i = 1, p^0 \ge 0}^K p^0_{i,j} \exp(\lambda_i) \exp(\tau_j) + \sum_{i = 1, p^0 < 0}^K \frac{p^0_{i,j}}{\exp(\lambda_i) \exp(\tau_j)} \right) \nonumber \\
    & \quad + \frac{p^0_{\phi,j}}{\exp(\tau_j) \exp(\eta)} + \left( \sum_{\substack{i = K+1, i \neq \phi, \\ p^0 \ge 0}}^{K + L} p^0_{i,j} \exp(\tau_j) + \sum_{\substack{i = K+1, i \neq \phi, \\ p^0 < 0}}^{K + L} \frac{p^0_{i,j}}{\exp(\tau_j)} \right) \nonumber \\
    & = y_j, \ (1 \le j \le K, \ p^0_{\phi,j} < 0) \label{eq:lambda2_n}
\end{align}
\begin{align}
    & \sum_{j=1, p^0 \ge 0}^K p_{\phi,j}^0 \exp(\tau_j) \exp(\eta) + \sum_{j=1, p^0 < 0}^K \frac{p_{\phi,j}^0}{\exp(\tau_j) \exp(\eta)} \nonumber \\
    & = \sum_{i=1, p^0 \ge 0}^K p_{i,\phi}^0 \frac{\exp(\lambda_i)}{\exp(\eta)} + \sum_{i=1, p^0 < 0}^K p_{i,\phi}^0 \frac{\exp(\eta)}{\exp(\lambda_i)}. \label{eq:lambda0}
\end{align}
The solutions for $\exp(\lambda_i)$, $\exp(\tau_j)$, and $\exp(\eta)$ are determined by solving the system of nonlinear equations formed by Eq. (\ref{eq:lambda1_p}) - (\ref{eq:lambda0}).
Substituting these values into Eq. (\ref{eq:p_a})-(\ref{eq:p_phir}) provides the estimates of $p_{i,j}$.

In this study, the ``root'' function of the ``optimize'' module in the SciPy library for Python was used to solve the system of equations numerically.

\paragraph{Acknowledgments}
The author expresses gratitude to Dr. Tomohiro Iwamoto (Nagoya Gakuin University) for providing relevant research materials and data that enhanced this study.
The author also thanks Dr. Ryutaro Nozaki (Kurume University) for his insightful advice on presenting the estimation results.
Furthermore, the author appreciates Dr. Mitsuru Morita (Aoyama Gakuin University) for his discussions on the research content, including estimation methods.
The author would like to thank Editage (www.editage.com) for English language editing.

The entire manuscript was edited and proofread in English using DeepL. After the authors reviewed the results, Editage conducted additional proofreading.

\end{document}